\documentstyle[aaspp4,amssym,psfig,epsfig,12pt]{article}

\lefthead{Fromerth, Melia, and Leahy}
\righthead{Iron 6.4 keV Emission at the Galactic Center}

\begin{document}
\centerline{Submitted to the Astrophysical Journal Letters}
\bigskip
\title{A Monte Carlo Study of the 6.4 keV Emission at the Galactic Center}

\author{Michael J. Fromerth\altaffilmark{1}}
\affil{Department of Physics, The University of Arizona, Tucson, AZ 85721}

\author{Fulvio Melia\altaffilmark{2}}
\affil{Department of Physics and Steward Observatory, The University of Arizona, Tucson, 
AZ 85721}
\altaffiltext{1}{NSF Graduate Fellow.}
\altaffiltext{2}{Sir Thomas Lyle Fellow and Miegunyah Fellow.}

\and

\author{Denis A. Leahy}
\affil{Department of Physics and Astronomy, University of Calgary, Calgary, Canada, T2N 1N4}

\begin{abstract}

Strong fluorescent Fe line emission at 6.4 keV has been observed from the Sgr~B2 
giant molecular cloud located in the Galactic Center region.  The large equivalent 
width of this line and the lack of an apparent illuminating nearby object indicate that 
a time-dependent source, currently in a low-activity state, is causing the fluorescent 
emission.  It has been suggested that this illuminator is the massive black hole
candidate, Sgr A*, whose X-ray luminosity has declined by an unprecedented six
orders of magnitude over the past $300$ years. We here report 
the results of our Monte Carlo simulations
for producing this line under a variety of source configurations and characteristics.
These indicate that the source may in fact be embedded within Sgr~B2, although external 
sources give a slightly better fit to the data.  The weakened distinction between
the internal and external illuminators is due in part to the instrument response
function, which accounts for an enhanced equivalent width of the line by folding
some of the continuum radiation in with the intrinsic line intensity.  We also point
out that although the spectrum may be largely produced by K$\alpha$ emission 
in cold gas, there is some evidence in the data to suggest the presence of warm 
($\sim 10^5$ K) emitting material near the cold cloud.

\end{abstract}


\section{INTRODUCTION}
The Galactic Center (GC) contains several constituents whose mutual interactions account
for a broad range of complex phenomena (e.g., Melia 1994).  Among these, the nonthermal radio source
Sgr A*, which appears to be coincident with the peak in the dark matter concentration
(e.g., Haller et al. 1996; Eckart \& Genzel 1997; Ghez et al. 1998), is of special 
interest since its properties suggest it is a massive ($2.6\pm0.2\times 10^6\;M_\odot$)
black hole.  In this context, the unusually strong iron fluorescent emission detected at 
6.4 keV near Sgr A*, particularly from the Sgr~B2 giant molecular cloud located some $90$ 
pc away, has raised some interesting questions regarding its origin and its possible diagnostic
value for understanding the high-energy behavior of the central engine. 

The GC is characterized by the presence of giant molecular clouds having mean volume 
densities of $\sim 10^4\ \mathrm{cm}^{-3}$ and gas temperatures on the order of 60~K 
(\cite{Lis94}).  X-ray observations of this region carried out by $ASCA$ 
(\cite{Koyama96}; \cite{Murakami00}) and $Beppo$SAX (\cite{Sidoli99}) have revealed 
a source of bright Fe fluorescent K$\alpha$ line radiation within Sgr~B2, whose radius 
is $\sim 20$ pc and whose total enclosed mass is $\sim 6 \times 10^6\ M_{\odot}$ (\cite{Lis94}).
This fluorescent emission has a very large equivalent width ($EW \approx 2-3$ keV) and a 
peak around 6.4 keV.  The surrounding continuum is quite flat, and shows strong absorption 
below $4.5$ keV and a sharp iron K$\alpha$ absorption feature at $7.1$ keV.

The large equivalent width is key to understanding the nature of the fluorescent 
emission (\cite{Sunyaev98}).  X-ray fluorescence is due to illumination of cloud material 
by an X-ray source.  For a steady source embedded within a gaseous cloud, an upper limit 
to the equivalent width is $\sim 1$ keV (e.g., \cite{Fabian77}; \cite{Vainshtein80}), 
rather insensitive to the parameters chosen.  Therefore, in the case of Sgr~B2, we are 
necessarily dealing with either a source external to the cloud or a time-dependent internal source whose 
flux has diminished, allowing the continuum to fade away relative to the line intensity.
An additional argument in favor of a time-dependent illuminator
is currently the absence of any bright X-ray source in the vicinity of Sgr~B2 
with sufficient luminosity to produce the observed fluorescence (e.g., \cite{Sidoli99}; 
\cite{Murakami00}); it seems that the illuminating object, whether internal or external 
to the cloud, must have undergone a decrease in X-ray flux to its current quiescent state.
Certainly, the SIGMA survey of this region (Vargas et al. 1996) has indicated that several
sources may remain difficult to detect due to their overall faintness and variable
behavior. 

It is tempting to invoke Sgr A* as the external illuminator of Sgr~B2, which would then
provide some measure of its recent variability at X-ray energies.  Motivated by the Sgr~B2 
spectral characteristics and the source morphology, which show that the emission is 
strongest on the side of the cloud closest to Sgr~A*, Murakami et al. (2000) suggested 
that we may be witnessing evidence of the black hole's activity some $300$ years ago 
(the light travel time from the nucleus to Sgr~B2).  These authors modeled a reflection 
spectrum using approximations for the photon transfer equations and obtained a decent 
fit to the observed spectrum.  However, it is important to note that this modeling is only 
a rough approximation, neglecting multiple Thomson scatterings and assuming isotropic 
reflection.  It is also limited by the fact that it does not explore the possibility
that the X-ray sources are internal to the cloud or that they are time-dependent, which
would provide important comparative results for fully testing the idea that Sgr A* was
the transient high-energy source illuminating the fluorescing region. 

There are several reasons for taking a skeptical view of Sgr A* as the X-ray illuminator.
First and foremost, this would require a change in its X-ray luminosity by a factor
of about $10^6$ in $300$ hundred years (from $L_x \approx 10^{39}$ erg s$^{-1}$, \cite{Sunyaev98}, 
to the currently observed value of $L_x \lesssim 10^{33}$ erg s$^{-1}$).  Secondly,
we are reminded of the fact that no other molecular cloud in the region, with the 
exception of the Sgr~A shell, displays any significant Fe K$\alpha$ emission 
(\cite{Koyama96}; \cite{Sidoli99}).  In the case of the Sgr~A shell, the K$\alpha$ 
line has a much smaller equivalent width ($\sim 100$ eV), easily explained by, e.g., 
a steady X-ray source embedded within the cloud.  It is therefore our intention in this
{\sl Letter} to explore alternative scenarios for the illumination of Sgr~B2,
using time-dependent X-ray sources both internal and external to the cloud, with
Monte Carlo methods to accurately model the spectrum.

\section{METHODOLOGY}

To simulate the X-ray spectrum formed under a wide range of conditions, we use a modified 
version of Leahy \& Creighton's (1983) \texttt{monte.f} code.  We have rewritten this
algorithm, originally designed to handle a time-steady photon transfer from internal
sources, to allow for blackbody and thermal bremsstrahlung continua, a time-dependence 
due to non-steady illuminators, and the presence of both internal and external X-ray sources.
A very important addition is that the escaping photons are now convolved with the $ASCA$ 
GIS response function and assigned to actual $ASCA$ GIS energy bins (rather than arbitrary 
logarithmic energy intervals), enabling a direct comparison with the actual data.

The photon propagation within the cloud, including the effects of electron scattering, 
absorption, and fluorescence, is handled as in \texttt{monte.f} (c.f. \cite{Leahy93} for 
a full description).  In particular, the bound-free absorption cross-sections are taken 
from Morrison \& McCammon (1983), the elemental abundances are from Dalgarno \& Layzer 
(1987), and the fluorescence yields are those given by Bambynek et al. (1972).

For blackbody and thermal bremsstrahlung spectra the photon distribution function 
$N(E)\,dE$ is too complex to allow direct Monte Carlo sampling.  Instead, we use 
rejection methods described in Press et al. (1992) to generate photons.  We choose 
a simpler function $f(E)\,dE$ (called the \emph{comparison function}) such that it lies 
above $N(E)\,dE$ everywhere within the relevant range of energies.  The comparison 
function is sampled using standard Monte Carlo techniques to determine an energy 
$E^{\prime}$.  If 
\begin{equation}
E_{min} \leq E^{\prime} \leq E_{max} \ ,
\label{eq:E_prime}
\end{equation}
where $E_{min}$ and $E_{max}$ are pre-defined lower and upper energy bounds, we 
multiply $f(E^{\prime})$ by a standard deviate $\xi$. When 
\begin{equation}
\xi f(E^{\prime}) \leq N(E^{\prime}) \ ,
\label{eq:photon_gen}
\end{equation}
we accept the result and a photon is generated with energy $E^{\prime}$.
If not, we reject the result and repeat the process until conditions (\ref{eq:E_prime}) 
and (\ref{eq:photon_gen}) are satisfied.

A blackbody spectrum of temperature $T$ has a normalized photon distribution function 
$N(x)\,dx = 15 x^3 dx / [\pi^4 (e^x - 1.0)]$, where $x \equiv E/T$ and both $E$ and $T$ 
are measured in keV.  To generate this spectrum, we choose the piece-wise continuous 
comparison function
\begin{equation}
f_{bb}(x) = \left\{ \begin{array}{ll} 0.15625 x \ , &  {\mathrm for\ } x < 2 \\ 0.3125 
\ , & {\mathrm for\ } 2 \leq x < 4 \\ 20/x^3 \ , & {\mathrm for\ } x \geq 4 \end{array} \right\} \ .
\label{eq:comp_bb}
\end{equation}
Note that we have normalized $f_{bb}(x)$ such that 
\begin{equation}
2\int_0^2 f_{bb}(x) dx = \int_2^4 f_{bb}(x) dx = \int_4^{\infty} f_{bb}(x) dx \ .
\label{eq:norm_bb}
\end{equation}

We approximate the thermal bremsstrahlung photon distribution function with cut-off temperature 
$T$ as $N(x)\,dx \sim x^{-1.2} e^{-x} dx$, with $x$ defined as above; our piece-wise continuous 
comparison function $f_{tb}(x)$ is then
\begin{equation}
f_{tb}(x) = \left\{ \begin{array}{ll} 0 \ , &  {\mathrm for\ } x < E_{min}/T \\ x^{-1.2} 
\ , & {\mathrm for\ } E_{min}/T \leq x < 1 \\ e^{-x} \ , & {\mathrm for\ } x > 1\end{array} \right\} \ .
\label{eq:comp_tb}
\end{equation}
In this case, the normalization is such that
\begin{equation}
\int_{E_{min}/T}^1 f_{tb}(x) dx = 5 e \left[ \left(\frac{T}{E_{min}}\right)^{0.2} -1 
\right] \ \int_1^{\infty} f_{tb}(x) dx \ .
\label{eq:norm_tb}
\end{equation}

To model time dependence, we assume that the source has undergone a period of intense emission 
that has diminished over a time interval much smaller than the light-crossing time of Sgr~B2.
In this limit, we create all photons at time $t = 0$ and track their travel time between 
interactions within the cloud.  Upon the last interaction, an additional time $(r + {\textbf r} 
\cdot {\textbf n})/c$, where ${\textbf r}$ is the location of the photon at last interaction 
and ${\textbf n}$ is the directional vector of the escaping photon, is added to each photon.
This factor represents the geometrical time lag for an escaping photon to reach the observer; 
similar lags appear in, e.g., reverberation mapping of active galactic nuclei (\cite{Blandford82}).
Escaping photons are then assigned to time bins of width $\Delta t = R/2c$, where $R$ is the 
cloud radius, with spectra being constructed for each of these time bins.

For simplicity, we assume that Sgr~B2 is a uniform spherical cloud; its optical depth and 
elemental abundances are allowed to vary, as well as the shape of the illuminating X-ray 
spectrum and the source location.  The flux density is also kept as a free parameter, since
fixing this would require further assumptions about the nature of the X-ray source, which
is not necessary for the modeling we describe here.

We consider both internal and external sources.  The former are assumed to lie at the center 
of the cloud and to emit isotropically, therefore maintaining spherical symmetry.  The escaping 
photons are binned only in time, as there is no need for angular bins due to the isotropy.
The external sources are assumed to lie at a distance much greater than the radius of the cloud, 
so that incoming photon distribution is roughly plane parallel.  The cloud is located with its 
center at the origin, and photons are incident from $+\hat{z}$-axis.  They enter the sphere at 
polar angle $\theta = \sin^{-1}(\xi)$, where $\xi$ is a random deviate chosen for each photon; 
the azimuthal angle $\phi$ is chosen randomly for each photon.  Each incident photon is assigned 
an initial time $(1 - \cos\theta)R/c$, representing the geometrical time lag for the photon 
to reach the sphere's surface.  As spherical symmetry has been broken, escaping photons are 
binned in $\theta$; we typically use four $\theta$ bins of width $\pi / 4$ each.
Azimuthal symmetry is retained, so no $\phi$ binning is necessary.

The modeled spectra are compared to the background-subtracted data of Murakami et al. (2000).
These observations were made using $ASCA$ during October 1993 and September 1994.
The escaping photons are convolved with the $ASCA$ GIS response function and binned according 
to the observed energy intervals.  Each model run consists of several million photons, and
the spectra are constructed for each time and $\theta$ bin.

\section{RESULTS}

In this section, we describe illustrative (reasonable) fits to the data.  Because there is a 
wide range of parameter values giving acceptable results and because the results are sensitive 
to, e.g., our selection of time and angular bin boundaries, it is not yet worthwhile attempting
to make a best fit, which is unlikely to be the ``unique'' solution.  In addition, our use of
a cold emitting gas appears to be lacking a key ingredient of the observed emission 
(c.f., Section~\ref{sec:Discussion}), making a best-fit analysis premature.

\subsection{Internal Sources}

An internal source must have a spectrum that drops off steeply at higher energies in order 
to explain the observed flat continuum.  This is because the absorption cross-section decreases 
at higher energies; there must be fewer incident photons at these energies in order for the 
escaping continuum to remain flat.  We find that blackbody and thermal bremsstrahlung sources, 
with their exponential cut-offs, provide decent fits to the data.

\begin{figure}
\figurenum{1}
\plottwo{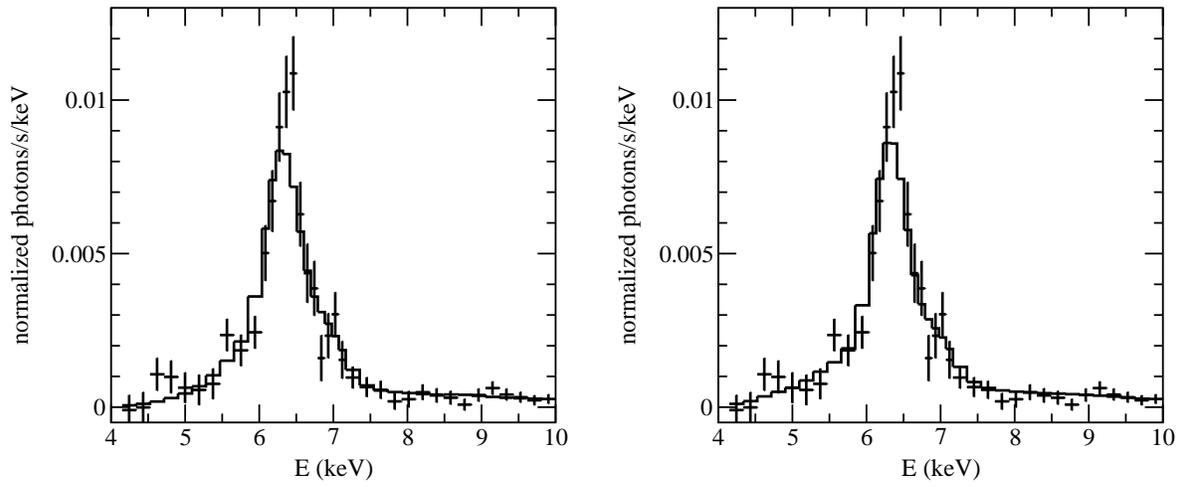}{fig2.eps}
\caption{Model fits (solid curves) to observed data (crosses) with time-dependent internal 
sources.  Left -- blackbody source.  Right -- thermal bremsstrahlung source. See text for
the assumed parameter values.}
\label{fig:internal}
\end{figure}

Figure~\ref{fig:internal} shows reasonable fits for blackbody (left) and thermal bremsstrahlung 
(right) internal sources.  In both cases, we used a cloud optical depth of $\tau = 0.2$ and 
elemental abundances enhanced to three times the local ISM values.  The blackbody and 
bremsstrahlung illuminating spectra have $T = 1$ keV and $3$ keV, respectively, and both have 
been allowed to evolve over time. The output spectra illustrated here are those that would
be observed at time $3R/c \leq t < 7R/2c$ (blackbody) and $5R/2c \leq t < 3R/c$ (bremsstrahlung).
Note that the first photons escape the sphere at $t = R/c$.  The fits lack 
an emission component at $\sim 6.46$ keV. For these models, we obtain $\chi_\nu^2 = 1.23$ 
(33 d.o.f.) for the blackbody and $\chi_\nu^2 = 1.17$ (33 d.o.f.) for the bremsstrahlung illuminators.

\subsection{External Sources}

Unlike the case of the internal sources, where the illuminator must apparently have a time-dependent
luminosity, we find that both time-steady and time-dependent external sources provide good fits to 
the data.  The best fits have incident spectra with $\alpha \approx -1$, where $N(E)dE \propto 
E^{\alpha} dE$, although there is much leeway here.  We note that for $x \equiv E/T\ll 1$, thermal 
bremsstrahlung spectra have $\alpha = -1.2$, so a hot, external bremsstrahlung source is 
certainly a possibility.

\begin{figure}
\figurenum{2}
\plottwo{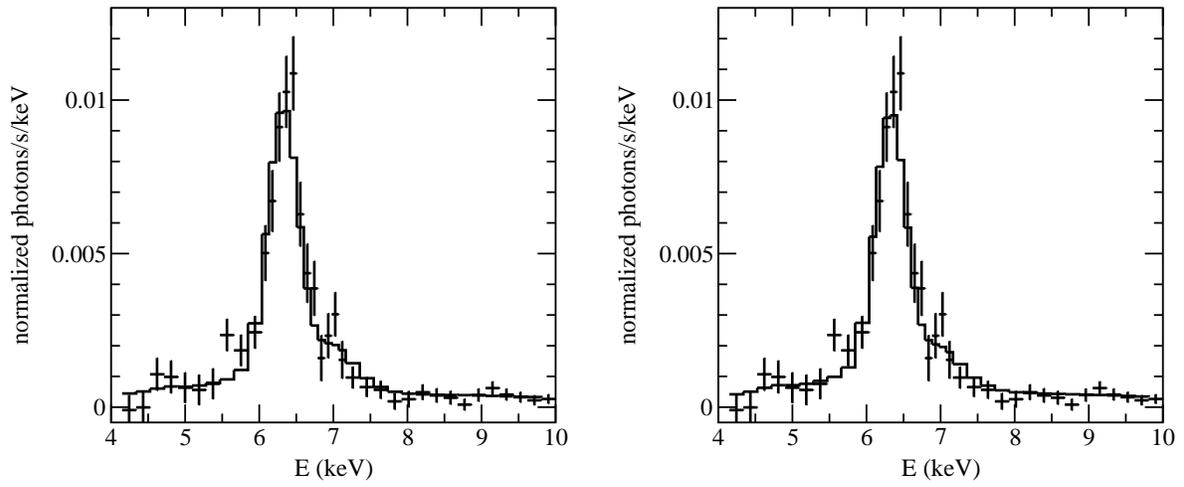}{fig4.eps}
\caption{Model fits (solid curves) to observed data (crosses) with external power law sources.  
Left -- time-steady source.  Right -- time-dependent source. See text for the assumed
parameter values}
\label{fig:external}
\end{figure}

Figure~\ref{fig:external} shows reasonable fits for time-steady (left) and time-dependent 
(right) external sources.  In both models, we used $\alpha = -1$, a cloud optical depth of 
$\tau = 0.4$, and elemental abundances enhanced to three times the local ISM values.
The spectra plotted are for $\pi/2 \leq \theta < 3 \pi / 4$, although the $\pi/4 \leq \theta 
< \pi/2$ spectra are nearly as good.  In the time-dependent model we are looking at the
spectrum formed during the period $R/c \leq t < 3R/2c$, which is early in the time-evolution 
and therefore (not surprisingly) quite similar to the time-steady case.  The fits appear to
be slightly better than those of the internal models, with $\chi_\nu^2 = 1.05$ (33 d.o.f.) 
for the steady source and $\chi_\nu^2 = 1.06$ (33 d.o.f.) for the time-dependent source.
Although not as apparent here, these spectra also lack an emission component at $\sim 6.46$ keV.

\section{DISCUSSION}
\label{sec:Discussion}

We have shown that the bulk of the iron fluorescent emission can be explained by either 
internal or external X-ray sources, though internal sources must be time-dependent; 
over time the line-emission becomes stronger relative to the continuum, giving an $EW$ 
approaching the observed value.  External sources may be time-steady or time-dependent, 
as it is possible to obtain a large $EW$ by angular selection effects.  However, a time-steady 
illuminator may be unlikely due to the current lack of an obvious source near Sgr B2,
with the possible exception of 1E 1743.1-2843, which lies only 63 pc (in projection) away
(e.g., Murakami et al. 2000).

The fact that internal sources are viable illuminators for the fluorescence in Sgr~B2 is
due in part to the limited energy resolution of the {\it ASCA} GIS detector.  Although the
equivalent width of the Fe 6.4 keV line is smaller for an internal source than for an
external illuminator, this situation changes when the calculated spectrum is convolved
with the detector response function.  It is evident from Figures 1 and 2 that
both internal and external sources give reasonable fits to the data.  Thus future
observations, e.g., with {\it XMM}, can in principle distinguish between these two types
of sources and help resolve the question of whether the iron line fluorescent emission
in Sgr~B2 does indeed indicate a past high-energy activity in Sgr A*. 

All the calculations reported here lack an emission component at $\sim 6.46$ keV.
It is possible that the appearance of this spectral feature is simply an artifact 
due to an incomplete background subtraction 
(since the Sgr~B2 background is very inhomogeneous).  If it is real, however, it 
may indicate emission from ``warm'' iron species (c.f., \cite{Nagase89}), in an ionization 
state \ion{Fe}{20} or greater.  Based on ionization calculations carried out with CLOUDY 
(\cite{Ferland96}), we find that these ``warm'' species exist only at temperatures on the 
order of $10^5$ K.  Gas at these high temperatures is unlikely to coexist with the cold 
cloud material, so one possibility is that it constitutes a transition layer between 
Sgr~B2 and the hot ($T \approx 10^6$ K) plasma surrounding it.  However, thermal conduction 
models seem to suggest that such a layer would be too thin to produce significant emission 
(\cite{Cowie77}).  Future observations may provide better clues to this puzzle.

\section{ACKNOWLEDGMENTS}

We thank Hiroshi Murakami for providing the $ASCA$ GIS data.  We also thank Rashid 
Sunyaev, Lara Sidoli and Sandro
Mereghetti for their helpful discussions.  This work was supported by an NSF Graduate 
Fellowship at the University of Arizona, by a Sir Thomas Lyle Fellowship and a 
Miegunyah Fellowship for distinguished overseas visitors at the University of Melbourne, 
and by NASA grants NAG5-8239 and NAG5-9205 at the University of Arizona.

\clearpage

\clearpage

\end{document}